\documentclass{article}

\usepackage{arxiv}

\usepackage[utf8]{inputenc} 
\usepackage[T1]{fontenc}    
\usepackage[
    colorlinks=false,
    linkcolor=blue,
    hidelinks,
    breaklinks]{hyperref}       
\usepackage{url}            
\usepackage{booktabs}       
\usepackage{amsfonts, amsmath}       
\usepackage{nicefrac}       
\usepackage{microtype}      
\usepackage{cleveref}       
\usepackage{graphicx}
\usepackage{doi}
\usepackage{caption, subcaption}
\usepackage[dvipsnames]{xcolor}

\title{Heracles: A HFO2 Ferroelectric Capacitor Compact Model for Efficient Circuit Simulations}

\date{}
\newif\ifuniqueAffiliation

\ifuniqueAffiliation 
\author{
}
\else
\usepackage{authblk}

\newbox{\orcid}\sbox{\orcid}{\includegraphics[scale=0.06]{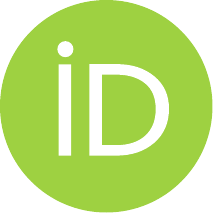}} 
\author[1]{%
	\href{https://orcid.org/0000-0003-0993-5593}{\usebox{\orcid}\hspace{1mm}Luca Fehlings\thanks{\texttt{Corresponding author: l.d.fehlings@rug.nl \\ This work has been submitted to the IEEE for possible publication. Copyright may be transferred without notice, after which this version may no longer be accessible.}}}%
}
\author[2]{%
	\href{https://orcid.org/0000-0002-0212-0298}{\usebox{\orcid}\hspace{1mm}Md Hanif Ali}%
}
\author[1]{%
	\href{https://orcid.org/0009-0002-6006-608X}{\usebox{\orcid}\hspace{1mm}Paolo Gibertini}%
}
\author[1,3]{%
	\href{}{\usebox{\orcid}\hspace{1mm}Egidio A. Gallicchio}%
}
\author[2]{%
	\href{https://orcid.org/0000-0002-1498-5993}{\usebox{\orcid}\hspace{1mm}Udayan Ganguly}%
}
\author[2]{%
	\href{https://orcid.org/0000-0002-0349-4857}{\usebox{\orcid}\hspace{1mm}Veeresh Deshpande}%
}
\author[1]{%
	\href{https://orcid.org/0000-0003-0479-6897}{\usebox{\orcid}\hspace{1mm}Erika Covi}%
}
\affil[1]{Zernike Institute for Advanced Materials \& Groningen Cognitive Systems and Materials Center (CogniGron)
	University of Groningen, 9747 AG Groningen, Netherlands}
\affil[2]{Department of Electrical Engineering,
Indian Institute of Technology Bombay, Mumbai 400076, India}
\affil[3]{Department of Applied Science and Technology, Politecnico di Torino, Turin, Italy.}
\fi

\renewcommand{\headeright}{}
\renewcommand{\undertitle}{}
\renewcommand{\shorttitle}{Heracles}

\hypersetup{
    pdftitle={Heracles: A HFO2 Ferroelectric Capacitor Compact Model for Efficient Circuit Simulations},
    pdfsubject={cs.ET, cond-mat.mtrl-sci, eess},
    pdfauthor={Luca Fehlings},
    pdfkeywords={memory device, hzo, hafnia, ferroelectric, compact model, tcad, spice, eda, circuit design},
}

\begin{document}
\maketitle

\begin{abstract}
    The growing use of ferroelectric-based technology, extending beyond conventional memory storage applications, necessitates the development of compact models that can be easily integrated into circuit simulation environments. These models assist  circuit designers in the design and the early assessment of the performance of their systems. The Heracles model is a physics-based compact model for circuit simulations in a SPICE environment for HfO$\mathrm{_2}$-based ferroelectric capacitors (FeCaps). The model has been calibrated based on experimental data obtained from HfO$\mathrm{_2}$-based FeCaps. A thermal model with an accurate description of the device parasitics is included to derive precise device characteristics based on first principles. The incorporation of statistical device data enables Monte Carlo analysis based on realistic distributions, thereby rendering the model particularly well-suited for design-technology co-optimization (DTCO). The model's efficacy is further demonstrated in circuit simulations using an integrated circuit with current programming, wherein partial switching of the ferroelectric polarization is observed. Finally, the model was benchmarked in an array simulation, reaching convergence in 1.8\,s with an array size of 100\,kb.
\end{abstract}

\section{Introduction}
\label{sec:introduction}
    Ferroelectric capacitors have the potential to be utilized in a multitude of applications that require a significant workload beyond that of memory storage \cite{ramaswamy-2023}. The advent of new paradigms, such as in-memory and analog computing, necessitates the application of the DTCO methodology. The integration of devices and circuits to reproduce sophisticated functionalities or accelerate computation while simultaneously optimizing power efficiency, density, and cost necessitates the development of compact models. These models must accurately describe the behavior of the devices and reliably converge when used with CMOS circuits in large arrays. In this particular context, the development of compact models that accurately capture the transient behavior of the FeCap, while also accounting for parasitics and variability, is imperative. This enables circuit designers to test the robustness of their system with a realistic behavior of the devices.
    
    A variety of models have already been proposed, with a particular emphasis on reproducing specific behaviors. These include quasi-static operation \cite{aziz-2016} and the dynamics of partial switching of the polarization \cite{tung-2021, yan-2023}. Typically empirical models such as multigrain Landau-Khalatnikov  \cite{pesic-2017, kim-2020}, Preisach \cite{ni-2018}, nucleation-limited switching (NLS) \cite{alessandri-2018} or domain wall movement models \cite{feng-2022} are employed in compact models. Other modeling approaches are based on specific and extensively measured characteristics such as the device impedance spectrum \cite{kim-2023}. However, a comprehensive model that can accurately describe all the behaviors exhibited by the FeCap is currently lacking. 
    
    In this study, we propose the Heracles device model, which is constructed on a physically plausible parameter set. This model facilitates collaboration among device engineers and circuit designers on a DTCO methodology to construct reliable and efficient memory circuits. Additionally, it enables the correlation of circuit reliability and performance metrics with physical parameters and concepts frequently employed in device development and reliability research.
    
     Heracles is a physics-based compact model of FeCaps based on Hf$\mathrm{_{0.5}}$Zr$\mathrm{_{0.5}}$O$\mathrm{_{2}}$ (HZO), validated against experimental data and incorporating a realistic distribution of the device-to-device variability. The model considers the intrinsic depletion of the electrodes and potential non-conductive or low-conductive interface layers, such as oxidized electrodes, and incorporates thermal models for both the polarization switching process and the leakage currents. The switching phenomenon then results from a complex process where the electric field experienced by the ferroelectric is affected by positive and negative feedback via the depletion and interface layer, leading to several characteristics observed in HfO$\mathrm{_2}$-based ferroelectric devices. Subsequently, the model is employed in circuit simulations, wherein it is subjected to current programming pulses. To assess its performance, Monte Carlo simulations are conducted and fitted against statistical device measurements. Finally, the model is tested for transient simulations on an array level to investigate its computational load and its convergence.
    
    \begin{figure}[!t]
        \centering
        \includegraphics[]{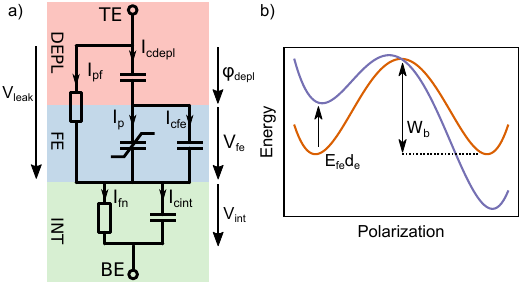}
        \caption{(a) Equivalent circuit of the model, with internal nodes for the interface and depletion layers. The voltage applied to the device is then distributed between the branches by the SPICE simulation. (b) Illustration of the energy profile based on Landau theory, where the ferroelectric switching process is modeled by a thermodynamic transition over the barrier $W_b$, modulated by the applied field (purple curve). }
        \label{fig:1}
    \end{figure}

\section{Physical device model}
        
    The compact model, which is available open source under the MIT license \cite{heracles_zenodo}, has been implemented in VerilogA and simulated using the Cadence Spectre SPICE simulator. The model is based on an equivalent circuit as depicted in Fig.~\ref{fig:1}(a) that models the behavior of three distinctive layers: The ferroelectric layer (FE), the electrode interface layer (INT) and the electrode depletion layer (DEPL). Each layer is composed of its respective currents: the depletion capacitance current $I_{cdepl}$ for the depletion layer, the current of the linear HZO capacitance $I_{cfe}$ and the ferroelectric polarization current $I_p$ for the ferroelectric layer, the leakage current $I_{pf}$ through the ferroelectric and depletion layers and the capacitive current $I_{cint}$ and leakage current $I_{fn}$ for the interface layer. An exhaustive list of all parameters with their corresponding fitting values can be found in Tab.~\ref{tab:params_nls}. 
    
    The intrinsic switching process of the ferroelectric layer is modeled by a non-equilibrium thermodynamic model, wherein the transition rates $k_{\downarrow/\uparrow}$ between the up and down polarization states are governed by a process with an energy barrier $W_b$:
    \begin{equation}
        k_{\downarrow/\uparrow} = \frac{k_B T}{h} \cdot \mathrm{exp} \left( \frac{(-W_b \pm W_e)}{k_B T} \right)   
        \label{eq:k}
    \end{equation}
    with Boltzmann constant $k_b$, Planck constant $h$ and temperature $T$. The applied energy $W_e$ then governs the switching probability as a function of the electric field in the ferroelectric layer $E_{fe}$:
    \begin{equation}
        W_e = (E_{fe} - E_{off}) \cdot d_e   
        \label{eq:we}
    \end{equation}
    with $d_e$ being the action distance of the electric field and $E_{off}$ the built-in bias field. This process is akin to the Landau free energy description applied to a ferroelectric (Fig.~\ref{fig:1}(b)), where an applied electric field shifts the energy landscape and thus the transmission probability from one state to the other. Accordingly, this results in the state variable $p$ (Eq.~\ref{eq:dpdt}) that describes the system's probability to be in the down polarization state:
    \begin{equation}
        \frac{dp}{dt} = k_\downarrow (1-p) - k_\uparrow p
        \label{eq:dpdt}
    \end{equation}
    Assuming that this switching process (or nucleation) of an elementary site does not directly interact with other elementary sites, we can macroscopically describe the proportion of domains in one of the polarization states to be equal to the occupation probability of that polarization state \cite{vopsaroiu-2010}. While this description is not compatible with domain wall movement models of conventional ferroelectrics, HfO$\mathrm{_2}$ domain wall hopping models suggest elementary sites with switching energy barriers around 1.38\,eV that switch cell by cell \cite{lee2020}, in accordance with our assumption of independently switching elementary sites. As $p$ is a probability assuming values between 0 and 1, we can calculate the polarization change, equivalent to the polarization displacement current $I_p$:
    \begin{equation}
         I_p = \frac{dP}{dt} = \frac{d(2 P_s p)}{dt}
         \label{eq:ip}
    \end{equation}
    where $P_s$ is the saturation polarization density of the ferroelectric layer.
    \begin{figure}[!t]
        \centering
        \includegraphics[]{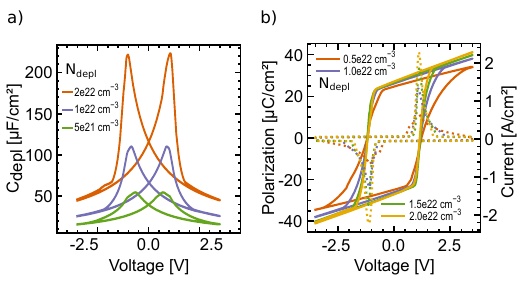}
        \caption{(a) Simulated small signal capacitance $C_{depl}$, a part of the total MFM capacitance (cf. Fig.~\ref{fig:3}(d)), for different carrier densities in the depletion layer. A hysteresis is observed due to the the independent modeling of the depletion layers for each polarization direction. (b) Simulated polarization hysteresis at 1 kHz for decreasing carrier densities in the depletion layer, where both the remanent polarization as well as the switching peak sharpness decrease as the carrier density in the electrode decreases. The dotted lines show the respective displacement currents.}
        \label{fig:2}
    \end{figure}
    Considering the full device, consisting of the ferroelectric layer, interface layers and electrodes, however, there is an interaction between these elementary sites, as the depolarization field acts as a feedback mechanism that modulates the electric field in the switching process proportional to the device polarization. Accordingly, at the macroscopic level, the elementary sites do not switch independently anymore due to the feedback via the global depolarization field.  Hence, an accurate description of the switching behavior needs to consider this depolarization field. We model this field via the capacitance of the dielectric and depletion interface layers. The dielectric capacitance of the ferroelectric ($C_{fe}$) and interface ($C_{int}$) layer are calculated as:
        \begin{equation}
        C_{fe/int} = \frac{\varepsilon_0 \varepsilon_{fe/int}}{t_{fe/int}}
        \label{eq:cfeint}
    \end{equation}
    where $\varepsilon_0$ is the vacuum permittivity, $\varepsilon_{fe/int}$ is the relative permittivity and $t_{fe/int}$ is the thickness of the respective layer. In particular, the depletion capacitance ($C_{depl \downarrow/\uparrow}$), which we model independently for both polarization directions \cite{kim-2023} similar to the polysilicon depletion model \cite{arora_1995}, has a pronounced influence on the switching process.
    \begin{equation}
        C_{depl \downarrow/\uparrow} = \frac{\varepsilon_{depl} q N_{depl \downarrow/\uparrow}}{\varepsilon_{fe} E_{fe} + Q_{fix,depl\downarrow/\uparrow}}
        \label{eq:cdepl_par}
    \end{equation} 
    with polarizability of the electrodes $\varepsilon_{depl}$, elementary charge $q$, carrier density in the electrode $N_{depl}$ and interface fixed charge $Q_{fix,depl\downarrow/\uparrow}$. Figure \ref{fig:2}(a) shows the depletion capacitance simulated for different carrier densities in the electrodes and exemplifies how modeling the capacitance independently for both polarization directions leads to the characteristic butterfly shape. The electric field originating from the depletion capacitance severely slows down the switching process due to the low series capacitance, as it can be seen by the widening of the polarization loop with decreasing carrier density in Fig.~\ref{fig:2}(b). This phenomenon can be severely impacted by charge screening due to trapped charges, modeled by the static charge $Q_{fix,depl}$, which partially screens the displacement originating from the polarization. The depletion capacitance in the electrodes are non-negligible due to the high displacement and thus charge density at the electrode interfaces \cite{black_1999}. This depletion effect can become even more impactful if one considers increased polarizability of the electrodes or decreased charge carrier densities in the electrodes due to the oxidation, interface effects or the thin-film nature of the electrodes. 

    \begin{table}[t]
        \centering
        \caption{Extracted model parameters}
        \setlength{\tabcolsep}{3pt}
        \begin{tabular}{|p{40pt}|p{75pt}|p{100pt}|}
        \hline
        Parameter& 
        Value& 
        Description\\
        \hline
        area & $625$\,µm² & Device area \\
        t$_{fe}$ & 9.8\,nm & FE thickness \\
        t$_{int}$ & 1\,nm & Int. thickness \\
        $\varepsilon_{fe}$ & 70 & FE relative permittivity \\
        $\varepsilon_{int}$ & 90 & Int. relative permittivity  \\
        $\varepsilon_{depl}$ & 3.6 & Depl. layer polarizability \\
        $W_{b}$ & 1.05\,eV & Switching energy barrier \\
        d$_{e}$ & 7.5\,nm & E-field action distance \\
        E$_{off}$ & 0.2\,MV/cm & Built-in bias field \\
        P$_s$ & 27\,µC/cm² & Saturation polarization \\
        N$_{depl}$ & $1.4 \cdot 10^{22}$\,cm$^{-3}$ & Depl. carrier density \\
        N$_{fe}$ & $1 \cdot 10^{18}$\,cm$^{-3}$ & FE Density of states \\
        $Q_{fix,depl\uparrow}$ & -9.45\,µC/cm² & Int. fixed charge \\
        $Q_{fix,depl\downarrow}$ & 9.45\,µC/cm² & Int. fixed charge \\
        m$_{eff,int}$ & 1 & Int. electron effective mass \\
        $\varphi_{b,int}$ & 0.65\,eV & Electrode/interface barrier \\
        $\varphi_{tr,fe}$ & 0.68\,eV & FE PF trap depth \\
        µ$_{fe}$ & $15 \cdot 10^{-4}$\,m$^2$V$^{-1}$s$^{-1}$ & FE charge carrier mobility \\
        \hline
        \end{tabular}
        \label{tab:params_nls}
    \end{table}
    \begin{figure}[!t]
        \centering
        \includegraphics[]{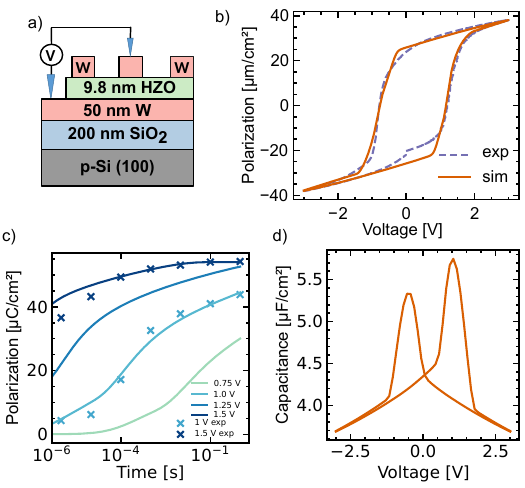}
        \caption{(a) Material stack of the metal-ferroelectric-metal (MFM) device from which the experimental data was gathered. (b) Hysteresis loop at 1 kHz for a woken-up device, resulting from the same parameters, showing that model generalizes the switching process well. (c) Simulated switching kinetics of the device and experimental data used for the parameter extraction. (d) Capacitance hysteresis of the whole device stack, simulated for a 100\,kHz, 30\,mV RMS sine.}
        \label{fig:3}
    \end{figure}
    The total depletion capacitance is then the sum of the depletion capacitance of the up and down state, weighted by the portion of the domains that are in the respective state:
    \begin{equation}
        C_{depl} = p \cdot C_{depl \downarrow} + (1-p) \cdot C_{depl \uparrow}
        \label{eq:cdepl}
    \end{equation}    
    The voltage $V_{fe}$ experienced by the ferroelectric layer, and governing the switching process, then depends on the voltage drop over the interface layer $V_{int}$ and the potential drop $\varphi_{depl}$ across the depletion capacitance $C_{depl}$:
    \begin{equation}
        \varphi_{depl} = \frac{2 P_s p - P_s + C_{fe} V_{fe}}{C_{depl}} 
        \label{eq:phidepl}
    \end{equation}
    which sums up to the total applied voltage $V_{app}$, the voltage between the top and the bottom electrodes:
    \begin{equation}
        V_{app} = V_{fe} + V_{int} + \varphi_{depl}
        \label{eq:vappl}
    \end{equation}
    A depolarization field then arises from the charges originating from the ferroelectric switching process which lead to a voltage over the interface and depletion capacitance. However, to accurately model the dynamics of the depolarization fields caused by the interface capacitance, we consider the leakage current ($J_{fn}$) through the interface layer via Fowler-Nordheim tunneling: 
     \begin{equation}
        J_{fn} = \frac{q^3}{8 \pi h q \phi_{b,int}} E_{int}^2 \mathrm{exp}\left( A \right)
    \end{equation}
    \begin{equation}
        A = -\frac{8\pi \sqrt{2m_0 m_{eff,int} } (q \phi_{b,int})^{3/2}}{3 h q E_{int}}
        \label{eq:jfn_a}
    \end{equation}
    with interface energy barrier height $\phi_{b,int}$, electron rest mass $m_0$ and electron effective mass in the interface $m_{eff,int}$.
    For the leakage current in the ferroelectric layer, which we consider the dominant contribution to the overall device leakage current, we implement a Poole-Frenkel conduction mechanism: 
    \begin{equation}
        J_{pf} = q \mu_{fe} N_{fe} E_{leak} \mathrm{exp} \left( B \right)
        \label{eq:jpf}
    \end{equation}
    \begin{equation}
        B = \frac{-q\left(\phi_{tr,fe} - \sqrt{\frac{qE_{leak}}{\pi \varepsilon_0\varepsilon_{fe}}}\right)}{kT}
    \end{equation}
    where $\mu_{fe}$ is the carrier mobility in the ferroelectric layer, $N_{fe}$ is the density of states in the conduction band and $\phi_{tr,fe}$ is the trap depth.
    As a result, the model allows to accurately describe several device behaviors that can be attributed to the contributions of macroscopic effects, such as the interface or depletion layer, with physical and experimentally verifiable parameters. Further, since the model is not relying on drawing from a statistical distribution or conditional branching to implement the ferroelectric switching process, the SPICE simulation is computationally efficient and well-posed, leading to reliable convergence. Ultimately, this allows the model to be used in Monte Carlo simulations reliably and efficiently.

\section{Experimental validation}
   
    To verify the model, we extract the model parameters (Table~\ref{tab:params_nls}) for an HZO FeCap with tungsten electrodes, as depicted in Fig.~\ref{fig:3}(a), with an area of 625\,µm². The fabrication details are described elsewhere \cite{ali-2024}, and all experimental measurements are performed after 1000 wake-up cycles at 3\,V, unless otherwise noted. The simulations were executed in the Cadence Spectre SPICE simulator at a at temperature of 21$^{\circ}$C and using the AHDL VerilogA compiler.
    
    With the extracted parameter set, we compare the resulting model simulations with the equivalent experimental characterization results from the device. In addition, we extract mean and standard deviation values for parameters that are a plausible physical source of the variability (Table~\ref{tab:params_mc}) and compare Monte Carlo simulations with an experimental device-to-device variability study of the switching kinetics at room temperature and 85°C.
    \begin{figure}[!t]
        \centering
        \includegraphics[]{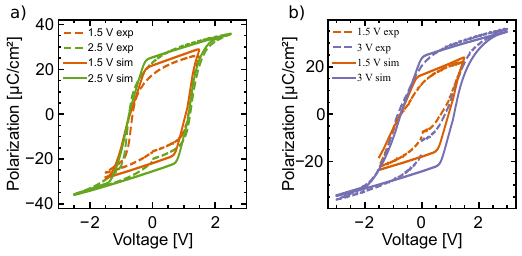}
        \caption{(a) Hysteresis at low voltages and 1\,kHz frequency that do not saturate the polarization and have a lower apparent coercive field due to the increased voltage ramp rate. (b) Hysteresis of a pristine device for different voltages, obtained by adjusting $N_{depl}$ to 7$\cdot$10$\mathrm{^{21}}$\,cm$^{-3}$, highlighting the increase of carrier density at the electrode interface as a possible contribution to the wake-up mechanism.}
        \label{fig:4}
    \end{figure}
    Using the extracted set of parameters, we compared the experimental polarization hysteresis curve (Fig.~\ref{fig:3}(b)) of a device after wake-up with the simulation and obtained a close match. Notably, the bias shift of the polarization hysteresis is accurately captured, however due to the asymmetry of the device the positive voltage switching process is captured more closely than the negative voltage portion. 
    The switching kinetics, over six orders of magnitude from 1\,µs to 1\,s, are modeled accurately as well, as evidenced by Fig.~\ref{fig:3}(c), following the experimental data for 1\,V and 1.5\,V. Further, the model also provides a capacitance hysteresis that resembles the characteristic butterfly shape (Fig.~\ref{fig:3}(d)) which originates from the electrode depletion capacitance. The capacitance hysteresis obtained also shows the typical asymmetry between the two capacitance peaks, which also reflects the voltage bias seen in the polarization hysteresis. In addition, the model replicates the hysteresis loop at lower voltages (Fig.~\ref{fig:4}(a)), where both the shift in coercive voltage and remanent polarization are approximated well. 
    As a first step to model the wake-up behavior of the device, with parameters adjusted for oxidation or degradation of the electrodes, lowering the depletion capacitance, the model is able to model the polarization hysteresis of the pristine device (Fig.~\ref{fig:4}(b)) and replicates both the change in coercive voltage as well as remanent polarization, also for sub-loops at a lower voltage of 1.5\,V. This suggests that the depletion capacitance, which we modified to fit the pristine polarization loops, is a plausible factor in the wake-up process. 
    \begin{figure}[!t]
        \centering
        \includegraphics[]{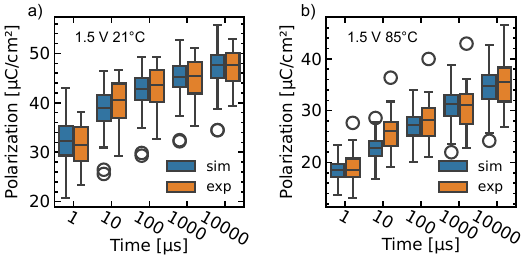}
        \caption{(a) Variability of the polarization switching kinetics at 21\,°C in the experiment and in the Monte Carlo simulation. (b) Switched polarization at 85\,°C, due to the elevated temperatures the polarization screening was modeled by a decrease in $d_e$ and an increase in $Q_{fix,depl}$ to account for increased charge trapping.}
        \label{fig:5}
    \end{figure}
    \begin{table}[!t]
        \centering
        \caption{Variability and temperature dependent parameters}
        \setlength{\tabcolsep}{3pt}
        \begin{tabular}{|p{40pt}|p{60pt}|p{60pt}|p{60pt}|}
        \hline
            Parameter & $\mu$ @ 21°C &  $\mu$ @ 85°C & $\sigma$ \\
            \hline
            t$_{int}$ & 1.5\,nm & 1.5\,nm &  0.22\,nm \\
            P$_{s}$ & 27\,µC/cm² & 27\,µC/cm² & 2.7\,µC/cm² \\
            N$_{depl}$ & $1.05 \cdot 10^{22}$\,cm$^{-3}$ & $1.05 \cdot 10^{22}$\,cm$^{-3}$ & $2.65 \cdot 10^{21}$\,cm$^{-3}$ \\
            Q$_{fix,depl}$ & 9.8\,µC/cm² & 27\,µC/cm² & 0 \\
            d$_e$ & 7.5\,nm & 4.5\,nm & 0 \\
        \hline
        \end{tabular}
        \label{tab:params_mc}
    \end{table}
    As a further step towards application in circuit design and DTCO methods, the model is validated against an experimental variability study of the switching kinetics experimentally obtained from 31 devices with Monte Carlo simulations (Fig.~\ref{fig:5}(a)) of 200 samples. The parameters used to model the variability (Tab.~\ref{tab:params_mc}), while not an exhaustive description of all possible variability sources, are able to model the median and quartiles of the experimental switching kinetics. 
    
    A further comparison was carried out on experimental switching kinetics at 85°C (Fig.~\ref{fig:5}(b)) where, with a parameter adjustment accounting for an increase in interface charges, the Monte Carlo simulation based on the thermal model also predicts both the switched polarization as well as the increasing spread of the polarization for longer time frames. The fact that the adjustment of the interface charge is needed to fit the higher temperature switching kinetics suggests that trapped charges, which become more pronounced at higher temperatures, play an important role in the transient behavior of the ferroelectric switching. Future developments of this model would therefore benefit from a model of the trapping process itself, instead of assigning a static interface charge, to accurately describe the dynamics of the interface charge and its temperature dependency.
    \begin{figure}[!t]
        \centering
        \includegraphics[]{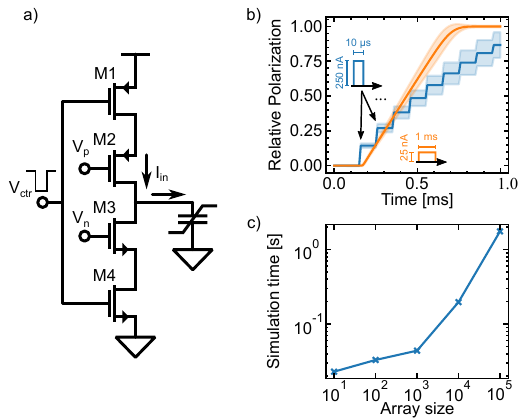}
        \caption{(a) Current programming circuit, with two transistors in saturation acting as current source (M2, M3) and switch (M1, M4). (b) Simulation of a 25 µm² FeCap programmed with a train of current pulses and a single pulse using the current programming circuit. The shaded area is the standard deviation obtained from the Monte Carlo simulation. (c) Simulation time, as a function of the array size, for a transient simulation of a single current pulse applied using the circuit the current programming circuit.}
        \label{fig:6}
    \end{figure}

\section{Circuit simulation}
    The availability of a stable compact model allows the exploration of new device applications that are hard to verify using standard test equipment such as semiconductor parameter analyzers and without a monolithic co-integration of FeCaps and CMOS circuits. One such case of study is the behavior of a FeCap programmed using a current source, which not feasible using discrete components due to the high parasitic capacitance. Here, the model is employed in a current programming circuit, as shown in Fig.~\ref{fig:6}(a). A current pulse is fed to the FeCap using two transistors in saturation (M1 and M4) with the biases $V_p$ and $V_n$ that set the positive and negative programming currents. The other two transistors (M2 and M3) are used to chop the current and thus either source or sink current. For the pulsed operation, a train of 250\,nA high and 10\,\textmu s wide current pulses is then applied to the FeCap, which is fully discharged to 0\,V after each pulse. For comparable continuous operation, a single 1\,ms pulse of 25\,nA, the average current of the train of pulses, is applied to the device.
    
    The Monte Carlo simulation in Fig.~\ref{fig:6}(b) shows the evolution of the polarization with the current pulses, including the standard deviation. The model can achieve partial polarization switching, even with current programming pulses, an important feature in applications that require analog computing or multilevel programming. Furthermore, it can predict the polarization switched by current input pulses, differentiating between different stimulating pulses. As evident from Fig.~\ref{fig:6}(b), the pulsed approach leads to a higher standard deviation in the programmed polarization compared to the continuous programming approach. This highlights that the model, due to its physical basis and well-posed description, is able both to generalize the device behavior well, allowing unconventional techniques such as current programming, as well as to converge reliably enough to allow Monte Carlo simulations and parameter explorations when designing memory macros. As such, the model allows the design of reliable FeCap-based circuits robust to device variation and can be an integral part in a design methodology for more reliable and performant memory macros. 
    
    Further, the model was tested in a 30\,\textmu s transient programming simulation using arrays of different sizes of the same programming circuit (Fig.~\ref{fig:6}(c)). In this case, the simulation reached convergence in 1.8\,s with an array size of 100\,kb, with the large array simulations possibly limited by the memory bandwidth in our simulation hardware. This is a promising result and very beneficial for the design of conventional memory arrays or analog crossbar arrays, where computationally heavy Monte Carlo and post-layout simulations of larger arrays give a crucial design advantage towards more reliable memory arrays.

\section{Conclusion}
    We developed a compact model of HZO FeCaps, implemented in VerilogA, with the specific objective of enabling circuit simulations for reliability, DTCO, and system-technology co-optimization (STCO). The model employs a comprehensive physics-based approach that accounts for a range of factors, including interface parasitic capacitances, leakage current, thermal effects, partial polarization switching, and device-to-device variations. The model has been calibrated using experimental data from several devices and enables Monte Carlo simulations based on a realistic distribution. 
    
    The functionality of the model is demonstrated in circuit simulation using a current programming circuit at both the single-device and array levels. Despite the inclusion of several intricate physical effects, the model converges and runs within a reasonable time frame for large device numbers, and for a wide range of parameters in Monte Carlo simulations. This signifies the model's readiness for practical circuit design. 
    
    Thanks to the separation of the model into components describing the ferroelectric, interface and depletion effects separately, the model is highly extensible and customizable. 
    
    Subsequent model enhancements are going to prioritize the development of more accurate leakage and charge trapping mechanisms with the goal of capturing the behavior of the device under endurance cycling. In conjunction with extensive physical characterization and TCAD modeling, which will provide a robust foundation for the physical parameters and their variability, this will enable circuit designs taking into account reliability and aging effects of the device. Additionally, together with a MOSFET model, Heracles can be extended to ferroelectric FETs (FeFETS). This extension will facilitate a more comprehensive understanding and modeling of the behavior and reliability of the HZO stack for highly scaled devices.

\section*{Acknowledgments}
    \small The Authors thank Dr. Fernando M. Quintana and Jorge Navarro Quijada for the fruitful discussions. This work was supported by the European Research Council (ERC) through the European's Union Horizon Europe Research and Innovation Programme under Grant Agreement No 101042585. Views and opinions expressed are however those of the authors only and do not necessarily reflect those of the European Union or the European Research Council. Neither the European Union nor the granting authority can be held responsible for them. The University of Groningen would like to acknowledge the financial support of the CogniGron research center and the Ubbo Emmius Funds. IIT Bombay acknowledges funding from DST and MeitY through NNETRA project, iHUB DivyaSampark - IIT Roorkee. Support of IITBNF staff is acknowleged.

\bibliographystyle{ieeetr}
\bibliography{references} 

\end{document}